\documentclass[12pt]{article}
\usepackage{graphicx}

\def\ignore#1{{}}

\newcounter{sxn}

\newcounter{axn}

\date{}

\newdimen\mybaselineskip
\newcommand{\beeq}{\begin{equation}}
\newcommand{\eneq}{\end{equation}}
\newcommand{\beqn}{\begin{eqnarray}}
\newcommand{\eeqn}{\end{eqnarray}}

\def\la{\raise.16ex\hbox{$\langle$}\lower.16ex\hbox{}  }
\def\ra{\, \raise.16ex\hbox{$\rangle$}\lower.16ex\hbox{} }

\def\psibar{ \psi \kern-.65em\raise.6em\hbox{$-$} \lower.6em\hbox{} }
\def\psibarb{ \psi \kern-.65em\raise.6em\hbox{$-$}  }

\begin{document}

\thispagestyle{empty}

\baselineskip=12pt



\vspace*{3.cm}

\begin{center}  
{\LARGE \bf Universality of Highly Damped Quasinormal Modes for Single Horizon Black Holes\footnote{Based on talk by G. Kunstatter given at Theory Canada 1, UBC, June, 2005.}}
\end{center}

\baselineskip=14pt

\vspace{3cm}
\begin{center}
{\bf  Ramin G. Daghigh\footnote{email: r.daghigh@uwinnipeg.ca} and Gabor Kunstatter\footnote{email: g.kunstatter@uwinnipeg.ca}}
\end{center}

\centerline{\small \it Department of Physics and Winnipeg Institute for Theoretical Physics,}
\centerline{\small \it  University of Winnipeg, Winnipeg, MB R3B 2E9, Canada. }
\vskip 1 cm

\vspace{3cm}
\begin{abstract}
\end{abstract}
It has been suggested that the highly damped quasinormal modes of black holes provide information about the microscopic quantum gravitational states underlying black hole entropy. This interpretation requires the form of the highly damped quasinormal mode frequency to be universally of the form: $\hbar\omega_R = \ln(l)kT_{BH}$, where $l$ is an integer, and $T_{BH}$ is the black hole temperature. We summarize the results of an analysis of the highly damped quasinormal modes for a large class of single horizon, asymptotically flat black holes.

\baselineskip=20pt plus 1pt minus 1pt

\newpage

\section{Introduction}
Black hole quasinormal modes (QNM's) are the natural vibrational modes of perturbations in the spacetime exterior to an event horizon. They are defined as solutions to the wave equation for the appropriate perturbation with boundary conditions that are ingoing at the horizon and outgoing at spatial infinity. The corresponding frequency spectrum is discrete and complex. The imaginary part of the frequency signals the presence of damping, a necessary consequence of boundary conditions that require energy to be carried away from the system.

The slowly damped QNM's (for gravitational perturbations) are relevant for astrophysical observations since they describe the frequency spectrum of the gravitational radiation that is expected to emerge from black hole formation during late times. The highly damped modes, i.e. the modes for which the damping rate goes to infinity, which are the subject of this paper, are unobservable. However, it has recently been suggested that the highly damped QNM's carry fundamental information about horizon dynamics and the microstates underlying black hole entropy. We begin by summarizing this proposed connection.

Numerical calculations of the QNM frequencies for Schwarzschild black holes in the early 90's revealed that in the limit of large damping, the frequency spectrum took the following form:
\beeq
\hbar\omega \to 2\pi i (n+{1\over2})kT_{BH} + (1.098612...)kT_{BH}~,
\eneq
where $T_{BH}$ is the Hawking temperature of the black hole.
The imaginary part became equally spaced (with $n$ large), whereas the real part of the frequency approached a constant. Note that since a Schwarzshild black hole is completely described by a single dimensionful parameter (the mass, or radius, or equivalently the temperature), it follows from dimensional grounds that the QNM frequency must be proportional to ${k\over \hbar} T_{BH}$. What is interesting about this spectrum is the fact that the constant of proportionality for the real part of the frequency approaches a universal value (i.e. independent of the angular momentum of the perturbation). Moreover, as Hod\cite{Hod} first noticed, the numerical value coincides to the given order with $\ln(3)$. (The fact that the coefficient was precisely $\ln(3)$ was later proved analytically by Motl\cite{Motl1}.) Hod went on to suggest a fascinating physical interpretation for this $\ln(3)$. Suppose, he said, that the limiting value of the real part of the highly damped QNM frequency was a fundamental vibrational mode associated with the dynamics of the event horizon itself. In this case, semi-classical arguments require the existence of states in the energy spectrum that are separated by the corresponding energy quantum:
\beeq
\Delta E_n = \hbar \omega \Delta n~,
\eneq
where $n$ is the integer labeling the states and $\Delta n=1$. In the large $n$ limit this expression can be integrated to yield:
\beeq
n=\int {dE\over \omega} = {1\over \ln(3)} \int {dE\over T_{BH}} ={1\over \ln(3)} S_{BH}~,
\label{first law}
\eneq
where $S_{BH}\propto Area/4$ is the Bekenstein-Hawking entropy\cite{Bekenstein-1,Hawking-1} of the black hole. Its appearance is a direct consequence of the first law of black hole thermodynamics. Equation (\ref{first law}) implies that the entropy/area is equally spaced in the semi-classical limit:
\beeq
S_{BH} = \ln(3) n = \ln(3^n)  ~.
\label{equally spaced}
\eneq
Amazingly, this form of the entropy is consistent with a statistical mechanic interpretation in terms of a black hole horizon made of $n$ fundamental elements of area, each with $3$ internal microstates.  This microscopic picture of black hole horizons was first conjectured by Bekenstein\cite{Bekenstein-2} and later Mukhanov\cite{bm}, who used it to argue for an equally spaced area spectrum of quantum black holes (although they assumed a binary structure for the area elements, so that the number of microstates was $2^n$). 

The above argument is of course highly conjectural. To have any hope of validity, it should apply in some form to any black hole event horizon, irrespective of the dynamics that lead to its formation. Specifically, it should apply to all asymptotically flat, single horizon black holes. This naturally raises the question of whether or not the coefficient of the real part of the frequency is generically $\ln(3)$. Motl and Neitzke\cite{Motl2} showed analytically that $\ln(3)$ is valid for higher dimensional Schwarzschild black holes, thereby verifying the conjecture in \cite{Gabor-prl}. 
More recently, Tamaki and Nomura\cite{tamaki} argued that the same coefficient was correct for 4-$d$ dilaton black holes, while
Kettner {\it et al}\cite{Gabor3} analyzed single horizon black holes in generic 2-$d$ dilaton gravity.
In a particularly elegant analysis, Das and Shankaranarayanan\cite{Das1} were able to study all single horizon black holes in 4 and higher dimensions with interesting results. Finally, the present authors \cite{Daghigh1} performed an analysis that included all single horizon, asymptotically flat black holes (including those considered in \cite{Gabor3} and \cite{Das1}) using the rigorous WKB formalism of Andersson and Howls\cite{Andersson}. The general and rigorous nature of this latter analysis gave significant insight into the source of the famous $\ln(3)$.  In particular, the numerical coefficient in the real part of the highly damped frequency is generically determined by the behaviour of coupling of the perturbation to the gravitational field near the origin, as expressed in tortoise coordinates. The $\ln(3)$ appears if and only if this coupling depends linearly on the tortoise coordinate near the origin. The question of universality seems to require an understanding of how this behaviour may, or may not, be connected to the dynamics of the horizon.

 In the next section, we set up the problem. Section 3 shows how Motl and Neitzke's monodromy calculation can be rigorously applied to generic single horizon, asymptotically flat black holes. The results, and their physical significance  for quantum gravity, are presented in the conclusions, along with a discussion of prospects for the future.

\section{QNM's For Generic Single Horizon Black Holes}

We wish to consider the general 2 dimensional scalar wave equation:
\beeq
\partial_\mu\left(\sqrt{-g}h(\phi)g^{\mu\nu}\partial_\nu \psi\right)=\sqrt{-g}V(\phi)\psi~,
\label{matter equation}
\eneq
where $g_{\mu\nu}$ is a two metric and $\phi$ is a scalar with respect to 2-$d$ coordinate transformations. 
Both the metric and dilaton are assumed static, so that one can find coordinates $(x,t)$ in which $\phi=\phi(x)$ and
 the metric takes the form
\beqn
ds^2&=&-f(x)dt^2+{1\over g(x)} dx^2~,\nonumber\\
&=& f(x)( -dt^2 + dz^2)~,
\label{metric}
\eeqn
where the second line expresses the metric in terms of the so-called ``tortoise'' coordinate $z$, defined by:
\beeq
dz = {dx\over F(x)}~,
\eneq
where $F(x)\equiv \sqrt{f(x)g(x)}$. The tortoise coordinate is distinguished by two features: the 2-metric is conformally Minkowskian, and $z\to -\infty$ logarithmically near an event horizon.

 The functions $f(x)$, $g(x)$, and $h(x)\equiv h[\phi(x)]$ are completely arbitrary at this stage, since we are making no assumptions about the gravitational dynamics or matter sources that give rise to this metric. By further restricting the coordinate system, it is possible to eliminate at most one of these functions, so the system is in fact completely specified by two arbitrary functions. 
   In order to restrict to single horizon black hole spacetimes we assume that $h(x)$ is monotonic and vanishes at $x=0$, which is a singular point in the spacetime. Moreover, we assume $f(x)$ and $g(x)$ have simple zeros at the same non-zero $x_h$, the horizon location. Their ratio $H(x)= {f(x)\over g(x)}$ is assumed to be a regular, nowhere vanishing, analytic function of $x$ \cite{Das1}.
The surface gravity of the corresponding black hole is given by:
\beeq
\kappa = {1\over 2} {dF\over dx}|_{x_h}~,
\eneq
and the associated Hawking temperature is generically given by:
\beeq
T_{BH}={\hbar \kappa \over 2\pi}~.
\eneq

The QNM's are obtained by looking for solutions to (\ref{matter equation}) that have the product form:
\beeq
\psi(x,t)= e^{-i\omega t} \Psi(x)~.
\eneq
If one defines a rescaled field $\overline{\Psi}= \sqrt{F}\Psi$, the wave equation in tortoise coordinates takes the simple form:
\beeq
\frac{d^2\overline{\Psi}}{dz^2}+\left( \omega^2-U_h(z) \right)\overline{\Psi}=0 ~,
\label{schrodinger}
\eneq
where 
\beeq
U_h\equiv {1\over2}{h''\over h}-{1\over4}\left({h'\over h}\right)^2+{F\over h}V(x)~,
\label{uh}
\eneq
and the prime here denotes differentiation with respect to $z$.  The potential $U_h$ goes to zero at both the horizon ($z\rightarrow -\infty$) and spatial infinity ($z\rightarrow \infty$).

The boundary conditions appropriate for QNM's are:
\beqn
\overline{\Psi}(z)&\to& e^{-i\omega z}\qquad \hbox{as $z\to -\infty (x\to x_h)$}\nonumber \\
  &\to& e^{+i\omega z}\qquad \hbox{as $z\to +\infty (x\to \infty)$}
\label{bc}
\eeqn
Our formalism applies to two distinct, but closely related (and overlapping) classes of black hole spacetimes.  First, one can consider (\ref{matter equation}), as in \cite{Gabor3}, to describe a scalar perturbation in two spacetime dimensions non-minimally coupled a black hole metric in generic 2-$d$ dilaton gravity.  In this case one can choose $\phi=x$ so that\cite{Gabor1}
\beeq
f(x)=g(x)= J(x)-2GM~,
\eneq
where $M$ is the mass of the black hole and $J(x)$ is determined by the dilaton potential, which is different for different theories.

Secondly, (\ref{matter equation})  describes the most general static, spherically symmetric metric in $d$ spacetime dimensions, with metric:
\beeq
d\hat{s}^2 = - {f}(r)dt^2+{1\over {g}(r)}dr^2 + r^2 d\Omega^{(n)}~,
\label{higher D}
\eneq 
where $n=d-2$ and $d\Omega^{(n)}$ is the line element on the unit $n$-sphere. This class of problems was considered by Das {\it et al}\cite{Das1}. 
 By dimensionally reducing the wave equation for a minimally coupled scalar field in this background and making the identifications $x=r$ and $h(x) = r^n $, one obtains precisely (\ref{matter equation}), with:
\beeq
V=r^n{l(l+n-1)\over r^2}~.
\eneq

\section{WKB/Monodromy Calculation}

The monodromy calculation proceeds by invoking the WKB approximation and calculating the change of the WKB phase along prescribed closed contours in the complex $x$-plane. The boundary conditions are imposed by relating the phase change along a contour that goes to infinity, where the solution is the prescribed outgoing wave, to the phase change around a contour very close to (and encircling) the horizon, where the form of the solution is known to be an ingoing wave (in tortoise coordinates). Demanding that the phase change calculated along the two contours be consistent gives an algebraic condition on the QNM frequency.

The trick is that while calculating the phase change along arbitrary contours on the complex plane is difficult, it is relatively easy if one sticks to a contour along which the WKB phase is purely real. These are the so-called anti-Stokes lines\cite{Andersson}. We therefore need to determine the structure of anti-Stokes lines in the complex $x$-plane. 

 Since we are interested in the highly damped QNM's where $|\omega_I| \rightarrow \infty$, the potential $U_h(z)$ is irrelevant in the region away from the origin.  In this region the anti-Stokes lines are the lines along which $\omega z$ is purely real.  A rough schematic behaviour of these lines are plotted in Fig. 1.  As one can see, we have two unbounded anti-Stokes lines which extend to infinity next to a bounded anti-Stokes line that loops around the event horizon.  Even if such unbounded anti-Stokes lines do not exist in one coordinate system, we can always generate such lines by a change of variable of the form $x \rightarrow \tilde{x}=x^q$, where $q$ is an integer greater than one.  Moving along the unbounded anti-Stokes lines in the clockwise direction and using the boundary condition at infinity, we find the monodromy around the contour A in Fig. 1 to be 
\beeq
\Psi \rightarrow e^{\pi \omega / \kappa} \chi_0 \Psi~,
\label{monodromy-A}
\eneq
where the $e^{\pi \omega / \kappa}$ is from moving along the solid line on which we have a plane wave solution of the form $e^{i \omega z}$, and $\chi_0$, to be determined later, is from moving along the dashed line in Fig. 1.

\begin{figure}[tb]
\begin{center}
\includegraphics[height=7cm]{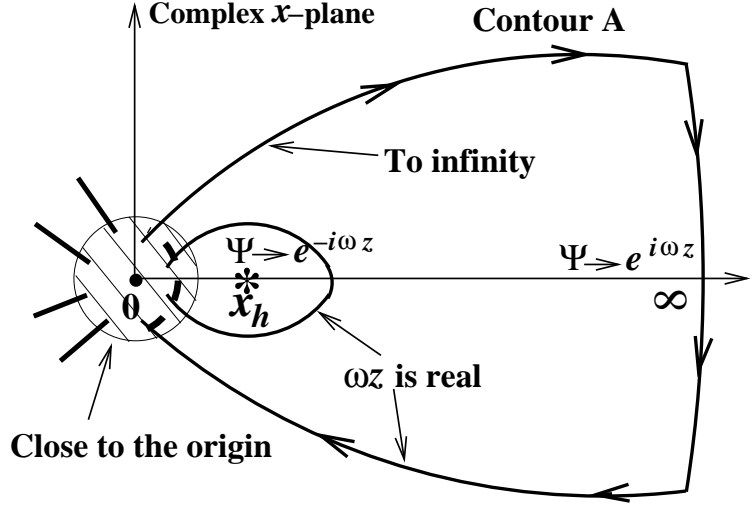}
\end{center}
\caption{Schematic of contours and Anti-stokes lines for generic monodromy calculation.}
\label{stokes}
\end{figure}

The monodromy around the same contour A can also be determined by observing that the only singularity inside this contour is at the event horizon. Thus the monodromy is the same as that of a small closed contour near the horizon. The boundary condition (\ref{bc}) requires this  monodromy to be:
\beeq
\Psi \rightarrow e^{-\pi \omega /\kappa}\Psi~.
\label{monodromy-h}
\eneq
Comparing Eqs. (\ref{monodromy-A}) and (\ref{monodromy-h}) gives the consistency condition
\beeq
e^{\pi \omega / \kappa} \chi_0 = e^{-\pi \omega /\kappa}~.
\label{WKB-condition}
\eneq
Once we determine $\chi_0$, we will be able to solve for the QNM frequency $\omega$.  To determine $\chi_0$, we need to know the behaviour of the solution near the origin where $U_h(z)$ diverges and therefore becomes relevant.  Assuming that $h(z) \rightarrow z^a$ as $z$ or $x\rightarrow 0$.  Then we have
\beeq
U_h(z) \rightarrow {{a(a-2)}\over {4z^2}}~,
\label{Uh-origin}
\eneq
close to the origin.  Thus, close to the origin, the relevant equation in tortoise coordinates is simply
\beeq
\frac{d^2\overline{\Psi}}{dz^2}+\left( \omega^2-{{a(a-2)}\over{4z^2}} \right)\overline{\Psi}=0 ~.
\label{schrod-origin}
\eneq
This equation can be solved exactly in terms of Bessel functions for generic $a$.  One interesting issue is that the rotation angle in the complex $x$-plane, which is the angle between the two unbounded anti-Stokes lines, is always correspond to a rotation by $3\pi$ in the tortoise coordinates.  Once we know the solution in this region, we can move along the dashed line in the clockwise direction and we can find $\chi_0$ which is
\beeq
\chi_0={1 \over -[1+2 \cos(\pi(a-1))]}~.
\label{chi-0}
\eneq
Inserting Eq. (\ref{chi-0}) into the consistency condition (\ref{WKB-condition}) will give us the condition
\beeq
e^{2 \pi \omega / \kappa} = -[1+2 \cos(\pi(a-1))]~.
\label{WKB-condition1}
\eneq
Using this equation we can get the QNM frequency
\beeq
\omega \mathop{\longrightarrow}_{|\omega_I| \rightarrow \infty} 2\pi i (n+{1\over 2})T_{BH}+ \ln[1+2\cos(\pi(a-1))]T_{BH}~.
\label{WKB-condition2}
\eneq

\section{Conclusion}

We have summarized a calculation that rigorously calculates the QNM frequencies for virtually all single horizon black holes, in any dimension. The coefficient of the real part of the QNM frequency generically is determined by the exponent $a$, which determines the rate at which the effective 2-$d$ coupling (i.e. $h(\phi)$ in Eq.(\ref{matter equation})) approaches zero at the origin, as expressed in tortoise coordinates. It is therefore at first glance difficult to see how this exponent is related to the dynamics of the horizon. Moreover, the form of the answer in the context of 2-$d$ dilaton gravity suggests that the coefficient of the real part is only the logarithm of an integer in exceptional cases. Although at first glance this also seems to be true for higher dimensional black holes, it is nonetheless encouraging that the $\ln(3)$ appears for all higher dimensional single horizon black holes, even those with non-trivial matter fields in which extra parameters in principle could affect the result. On the other hand, these simple and elegant results do not seem to apply to multi-horizon black holes \cite{Motl2, Andersson, multi-h}. These issues are currently under investigation.

\section{Acknowledgments}
We are grateful to Joey Medved and Joanne Kettner for their collaboration in an earlier part of this work. We also thank, Saurya Das, Brian Dolan and  S. Shankaranarayanan for useful discussions.  This research was supported in part by the Natural Sciences and Engineering Research Council of Canada.


\def\jnl#1#2#3#4{{#1}{\bf #2} (#4) #3}

\def\Zphys{{\em Z.\ Phys.} }
\def\jssc{{\em J.\ Solid State Chem.\ }}
\def\jpsJ{{\em J.\ Phys.\ Soc.\ Japan }}
\def\ptps{{\em Prog.\ Theoret.\ Phys.\ Suppl.\ }}
\def\PTP{{\em Prog.\ Theoret.\ Phys.\  }}
\def\LNC{{\em Lett.\ Nuovo.\ Cim.\  }}

\def\JMP{{\em J. Math.\ Phys.} }
\def\NPB{{\em Nucl.\ Phys.} B}
\def\NP{{\em Nucl.\ Phys.} }
\def\PLB{{\em Phys.\ Lett.} B}
\def\PL{{\em Phys.\ Lett.} }
\def\PRL{\em Phys.\ Rev.\ Lett. }
\def\PRB{{\em Phys.\ Rev.} B}
\def\PRD{{\em Phys.\ Rev.} D}
\def\PR{{\em Phys.\ Rev.} }
\def\PRe{{\em Phys.\ Rep.} }
\def\AP{{\em Ann.\ Phys.\ (N.Y.)} }
\def\RMP{{\em Rev.\ Mod.\ Phys.} }
\def\ZPC{{\em Z.\ Phys.} C}
\def\SCI{\em Science}
\def\CMP{\em Comm.\ Math.\ Phys. }
\def\MPLA{{\em Mod.\ Phys.\ Lett.} A}
\def\IJMPB{{\em Int.\ J.\ Mod.\ Phys.} B}
\def\cmp{{\em Com.\ Math.\ Phys.}}
\def\JPA{{\em J.\  Phys.} A}
\def\CQG{\em Class.\ Quant.\ Grav.~}
\def\ATMP{\em Adv.\ Theoret.\ Math.\ Phys.~}
\def\PRSA{{\em Proc.\ Roy.\ Soc.} A }
\def\ibid{{\em ibid.} }
\vskip 1cm

\leftline{\bf References}

\renewenvironment{thebibliography}[1]
        {\begin{list}{[$\,$\arabic{enumi}$\,$]}  
        {\usecounter{enumi}\setlength{\parsep}{0pt}
         \setlength{\itemsep}{0pt}  \renewcommand{\baselinestretch}{1.2}
         \settowidth
        {\labelwidth}{#1 ~ ~}\sloppy}}{\end{list}}


\end{document}